\documentclass[lettersize,journal]{IEEEtran}

\usepackage{cite}
\usepackage{amsfonts,amsmath,amssymb}
\usepackage[fleqn]{mathtools}
\usepackage{acronym}
\usepackage{tikz}
\usepackage{subfigure}
\usetikzlibrary{arrows.meta}
\usepackage{algorithm}
\usepackage[noend]{algpseudocode}
\usepackage{courier}
\usepackage[inline]{enumitem}
\usepackage{microtype}

\definecolor{matlab_blue}{rgb}{0.00,0.45,0.74}
\definecolor{matlab_red}{rgb}{0.85,0.33,0.10}
\definecolor{matlab_orange}{rgb}{0.87,0.49,0.00}

\newcommand{\blueline}{\tikz[baseline]{\draw[blue,solid,line width = 1.0pt](0,0.8mm) -- (5.3mm,0.8mm)}}
\newcommand{\redline}{\tikz[baseline]{\draw[red,solid,line width = 1.0pt](0,0.8mm) -- (5.3mm,0.8mm)}}
\newcommand{\bluetriangle}{\tikz[baseline]{\draw[blue,solid,line width = 0.5pt] 
(0.0mm,1.6mm) -- (2.0mm,1.6mm) --  (1.0mm,0.0mm) -- (0.0mm,1.6mm)}}
\newcommand{\blackcross}{\tikz[baseline]{\draw[black,solid,line width = 0.5pt] 
(0.0mm,0.0mm) -- (1.6mm,1.6mm) (0.0mm,1.6mm) -- (1.6mm,0.0mm)}}
\newcommand{\redplus}{\tikz[baseline]{\draw[red,solid,line width = 0.5pt]
(0mm,0.8mm) -- (1.6mm,0.8mm)  (0.8mm,1.6mm) -- (0.8mm,0mm) }}

\newcommand{\blacksquareline}{\tikz[baseline]{\draw[black,solid,line width = 1pt] (0.0mm,0mm) -- (2.0mm,0mm) --  (2.0mm,2.0mm) -- (0.0mm,2.0mm) -- (0.0mm,0mm)  (-2.0mm,1mm) -- (4.0mm,1mm)}}
\newcommand{\redtriangleline}{\tikz[baseline]{\draw[matlab_red,solid,line width = 1pt] (0mm,0mm) -- (1.5mm,2.0mm) --  (-1.5mm,2.0mm) -- (0.0mm,0.0mm);
\draw[matlab_red,dashed,line width = 1pt] (-3.5mm,1mm) -- (3.5mm,1mm)}}

\algrenewcommand\algorithmicindent{1.0em}

\makeatletter
\let\OldStatex\Statex
\renewcommand{\Statex}[1][3]{%
  \setlength\@tempdima{\algorithmicindent}%
  \OldStatex\hskip\dimexpr#1\@tempdima\relax}
\makeatother

\newcommand{\N}[0]{\ensuremath{\mathcal{N}}}
\newcommand{\atan}{ \textrm{atan}  }
\newcommand{\argmin}{ \textrm{arg\,min}  }
\newcommand{\imagunit}{\mathrm{j}}

\renewcommand{\vec}[1]{\ensuremath{{\mathbf{#1}}}}
\newcommand{\vecsymbol}[1]{\ensuremath{\boldsymbol{#1}}}

\newcommand{\vb}[0]{\vec{b}}
\newcommand{\vp}[0]{\vec{p}}
\newcommand{\vu}[0]{\vec{u}}
\newcommand{\vv}[0]{\vec{v}}

\newcommand{\vz}[0]{\vec{z}}
\newcommand{\vh}[0]{\vec{h}}

\newcommand{\vx}[0]{\vec{x}}
\newcommand{\vA}[0]{\vec{A}}
\newcommand{\vI}[0]{\vec{I}}
\newcommand{\vR}[0]{\vec{R}}
\newcommand{\vH}[0]{\vec{H}}

\newcommand{\vmu}[0]{\vecsymbol{\mu}}
\newcommand{\vnu}[0]{\vecsymbol{\nu}}

\newcommand{\cH}[0]{\mathcal{H}}
\acrodef{5G}[5G]{fifth generation}
\acrodef{6G}[6G]{sixth generation}
\acrodef{mmWave}[mmWave]{millimeter wave}
\acrodef{UE}[UE]{user equipment}
\acrodef{BS}[BS]{base station}
\acrodef{PRS}[PRS]{positioning reference signal}
\acrodef{LOS}[LoS]{line-of-sight}
\acrodef{NLOS}[NLoS]{non-line-of-sight}
\acrodef{TX}[TX]{transmitter}
\acrodef{RX}[RX]{receiver}
\acrodef{SLAM}[SLAM]{simultaneous localization and mapping}
\acrodef{TOA}[ToA]{time-of-arrival}
\acrodef{AOA}[AoA]{angle-of-arrival}
\acrodef{AOD}[AoD]{angle-of-departure}
\acrodef{BIC}[BIC]{Bayesian information criterion}
\acrodef{RANSAC}[RANSAC]{random sample consensus }
\acrodef{SCS}[SCS]{subcarrier spacing}
\acrodef{OFDM}[OFDM]{orthogonal frequency-division multiplexing}
\acrodef{RMSE}[RMSE]{root mean squared error}
\acrodef{TP}[TP]{true positive}
\acrodef{TPR}[TPR]{true positive rate}
\acrodef{TN}[TN]{true negative}
\acrodef{TNR}[TNR]{true negative rate}
\acrodef{MCS}[MCS]{Monte Carlo simulation}
\acrodef{CDF}[CDF]{cumulative distribution function}
\acrodef{mmWave}[mmWave]{millimeter-wave}

\begin{document}

\bstctlcite{IEEEexample:BSTcontrol}

\title{Robust Snapshot Radio SLAM}

\author{\IEEEauthorblockN{
Ossi Kaltiokallio,
Elizaveta Rastorgueva-Foi,
Jukka Talvitie,
Yu Ge,
Henk Wymeersch,
and Mikko Valkama
}                                     
\thanks{O. Kaltiokallio, E. Rastorgueva-Foi, J. Talvitie, and M. Valkama are with Tampere University, Finland.}
\thanks{Y. Ge and H. Wymeersch are with Chalmers University of Technology, Sweden. 
}
\thanks{This work has been supported by the Academy of Finland (grants \#338224, \#345654, \#352754 and \#359095), by Business Finland (6G-ISAC project), and by  the Vinnova B5GPOS Project under Grant 2022-01640.}
\vspace{-8mm}
}



\maketitle

\begin{abstract}
The intrinsic geometric connections between \ac{mmWave} signals and the propagation environment can be leveraged for \ac{SLAM} in 5G and beyond networks. However, estimated channel parameters that are mismatched to the utilized geometric model can cause the \ac{SLAM} solution to degrade. In this paper, we propose a robust snapshot radio \ac{SLAM} algorithm for mixed \ac{LOS} and \ac{NLOS} environments that can estimate the unknown \ac{UE} state, map of the environment as well as the presence of the \ac{LOS} path. The proposed method can accurately detect outliers and the \ac{LOS} path, enabling robust estimation in both \ac{LOS} and \ac{NLOS} conditions. The proposed method is validated using 60\,GHz experimental data, indicating superior performance compared to the state-of-the-art. 
\end{abstract}
\vspace{-4mm}
\begin{IEEEkeywords}
5G, 6G, mmWave, simultaneous localization and mapping, robust estimation
\end{IEEEkeywords}
\vspace{-4mm}
\section{Introduction}
The development of evolving 5G and future 6G networks provides not only new opportunities for improving the quality and robustness of communications, but also for facilitating high-accuracy localization and sensing \cite{BehravanVTM2023,delima2021}. This is enabled by increased temporal and angular resolution due to the use of higher frequency bands and thus larger bandwidths combined with larger antenna arrays. Extracting timely and accurate location and situational awareness is, in general, a critical asset in various applications such as vehicular systems and industrial mobile work machines \cite{LiuTCDS2022,BarnetoTVT2022}.

Localization and sensing using downlink/uplink signals between the \ac{BS} and \acf{UE} is often a two-stage process \cite{shahmansoori2018}, in which the channel parameters, that is, the \ac{AOA}, \ac{AOD} and \ac{TOA} of the resolvable propagation paths are first estimated. Then, in the second stage, the channel parameters are used for localization and sensing. The resolvable \acf{NLOS} paths provide information not only about the \ac{UE} position, but also regarding the incidence point of single-bounce \ac{NLOS} paths. While these incidence points are unknown, the cardinality of the channel parameters can outweigh the unknowns, enabling bistatic radio \acf{SLAM} which aims to estimate the unknown \ac{UE} state and landmark locations using the channel estimates and the known \ac{BS} state \cite{wymeersch2018}.

The bistatic \ac{SLAM} problem has gained widespread attention in the 5G/6G research field since it improves localization accuracy \cite{shahmansoori2018}, enables localizing the \ac{UE} in the absence of \acf{LOS} \cite{wen2021}, and reduces the dependence on infrastructure thus enabling localization even with a single \ac{BS} \cite{wymeersch2018}. To this end, two mainstream approaches exist in literature:
\begin{enumerate*}[label=(\roman*)]
  \item \emph{filtering-based solutions} that recursively estimate the joint state of the \ac{UE} and map by processing the observations sequentially over time;
  \item \emph{snapshot approaches} that solve the \ac{SLAM} problem for a single \ac{UE} location without any prior information nor kinematic models.
\end{enumerate*}
The main benefit of filtering approaches is that higher localization accuracy is expected and state-of-the-art filters \cite{ge2022,kim2020,kaltiokallio2022} can inherently deal with many of the challenges faced in bistatic \ac{SLAM} including false detections generated by clutter and multi-bounce propagation paths. In contrast, snapshot \ac{SLAM} is fundamentally important as it serves as a baseline for what can be done with radio signals alone \cite{shahmansoori2018,wen2021,ge2023,nazari2023} and the solution can be used to initialize and/or as input to filtering approaches. A major limitation of existing snapshot approaches is that they typically overlook outlier measurements that for example originate from multi-bounce propagation paths and that have a mismatch to the utilized geometric model
\cite{shahmansoori2018,wen2021,ge2023,nazari2023}. As a consequence, outliers will cause localization performance degradation.

In this paper, we address the limitations of the state-of-the-art in \cite{shahmansoori2018,wen2021,ge2023,nazari2023} and develop a novel snapshot \ac{SLAM} method to estimate the \ac{UE} state and map of the propagation environment using a single downlink transmission from one \ac{BS}, that is robust against outlier measurements. The contributions can be summarized as follows:
\begin{itemize}
    \item We propose a low-complexity processing pipeline for solving the snapshot \ac{SLAM} problem using measurements that are corrupted by outliers\textemdash defined as measurements not originating from \ac{LOS} or single-bounce \ac{NLOS} paths;
    \item We present a method to detect the \ac{LOS} path and solve the \ac{SLAM} problem in mixed \ac{LOS}/\ac{NLOS} conditions, as opposed to \cite{wen2021,ge2023} that always assume \ac{NLOS}; 
    \item Using experimental data with off-the-shelf 60\,GHz \ac{mmWave} MIMO radios, we demonstrate that the developed algorithm can accurately recover the inliers resulting in superior accuracy with respect to the state-of-the-art benchmark algorithms;
    \item We provide the channel estimates for the experimental data as well as Matlab code that runs the presented method. 
\end{itemize}

\section{Problem Formulation and Models}
 
Considering the experimental setup in Section \ref{sec:results} and to simplify the presentation, we focus on describing the methods in the 2D/azimuth domain but extension to 3D is possible (conceptually similar to, e.g., \cite{wen2021}). The \ac{BS} state is represented by the position $\vp_\textrm{BS} \in \mathbb{R}^2$ and orientation $\alpha_\textrm{BS}\in [-\pi,\pi]$. The $i$th landmark is represented by position $\vp_i \in \mathbb{R}^2$ which depicts the interaction point of the $i$th single-bounce propagation path. The \ac{UE} state, $\vx$, is described using the position $\vp_\textrm{UE} \in \mathbb{R}^2$, orientation $\alpha_\textrm{UE}$ and clock bias $b_\textrm{UE}$.

\subsection{System Model}

\begin{figure}[!t]
\centering
     \resizebox{88mm}{47mm}{
\begin{tikzpicture}

\draw[->,dashed,gray] (0,1) -- (0,4);
\draw[->,dashed,gray] (0,1) -- (3,1);
\draw[black,line width=0.3mm,->] (0,1) -- (2.9613,1.4805);
\draw[black,line width=0.3mm,->] (0,1) -- (-0.4805,3.9613);
\node[rotate=15,xshift=2mm,yshift=0mm] at (2.92,1.70) {$x''$};
\node[rotate=15,xshift=0mm,yshift=2mm] at (-0.70,3.92) {$y''$};

\node[xshift=2mm,yshift=2mm,gray] at (8,3) {$0$};
\node[right,xshift=0mm,yshift=2.5mm,gray] at (6,5) {$\tfrac{\pi}{2}$};
\node[left,xshift=0mm,yshift=2mm,gray] at (4,3) {$\pi$};
\node[left,xshift=0mm,yshift=-2mm,gray] at (4,3) {-$\pi$};
\node[left,xshift=0mm,yshift=-2.5mm,gray] at (6,1) {-$\tfrac{\pi}{2}$};

\draw[->,dashed,gray] (3,3) -- (9,3);
\draw[->,dashed,gray] (6,0.5) -- (6,5.5);
\node[xshift=2mm,yshift=0mm,gray] at (9,3) {$x$};
\node[xshift=0mm,yshift=2mm,gray] at (6,5.5) {$y$};
\draw[black,line width=0.3mm,->] (6,3) -- (6.85,0.65);  
\draw[black,line width=0.3mm,->] (6,3) -- (8.82,4.02); 
\node[rotate=20,xshift=0mm,yshift=-2mm] at (6.85,0.65) {$x'$};
\node[rotate=20,xshift=2mm,yshift=0mm] at (8.82,4.02) {$y'$};

\node[fill,draw,circle,color=gray,minimum size=1mm] at (6,3) (a) {};
\node[fill,draw,circle,color=gray,minimum size=1mm] at (0,1) (b) {};
\node[fill,draw,circle,color=gray,minimum size=1mm] at (1,5) (c) {};

\draw[blue, line width=0.3mm,-{Latex[length=3mm,width=3mm]}] (6,3) -- (0,1);
\draw[blue, line width=0.3mm,-{Latex[length=3mm,width=3mm]}] (6,3) -- (1,5);
\draw[blue, line width=0.3mm,-{Latex[length=3mm,width=3mm]}] (1,5) -- (0,1);

\draw[gray,dashed] (8,3) arc [start angle=0, delta angle=360, radius=2];
\draw[black] (6.5,3) arc [start angle=0, delta angle=-70, radius=0.5];
\draw[black] (2,1) arc [start angle=0, delta angle=9.2174, radius=2];
\draw[blue] (6.3420,    2.0603) arc [start angle=290, delta angle=-131.8014, radius=1];
\draw[blue] (6.4275,    1.8254) arc [start angle=290, delta angle=-91.5651, radius=1.25];
\draw[blue] (1.9742,1.3204) arc [start angle=9.2174, delta angle=9.2174, radius=2];
\draw[blue] (1.7274,1.2803) arc [start angle=9.2174, delta angle=66.7463, radius=1.75];
\draw[] (a) node[right,xshift=0mm,yshift=4mm] {$\vp_\textrm{BS}$};
\draw[] (b) node[right,xshift=0mm,yshift=-3mm] {$\vp_\textrm{UE}$};
\draw[] (c) node[right,xshift=0mm,yshift=3mm] {$\vp_i$};

\node[right,xshift=-0.5mm,yshift=1.75mm] at (2,    1) {$\alpha_\textrm{UE}$};
\node[right,xshift=-1.5mm,yshift=-4.00mm] at (6.5,3) {$\alpha_\textrm{BS}$};
\node[right,xshift=-15mm,yshift=-0.50mm] at (6.4275,    1.8254) {$\phi_0$};
\node[right,xshift=-0.5mm,yshift=4.5mm,rotate=10] at (2,    1) {$\theta_0$};
\node[rotate=18.4349,yshift=2mm] at (3,2) {$d_0\vu_0$};
\node[right,xshift=-20mm,yshift=14.00mm] at (6.4275,    1.8254) {$\phi_i$};
\node[right,xshift=-8mm,yshift=7mm,rotate=54.5] at (2,    1.5) {$\theta_i$};
\node[rotate=-21.8014,yshift=2mm] at (3.5,4) {$d_{i}\gamma_i\vu_i$};
\node[rotate=75.9638,yshift=2mm] at (0.6,3.5) {$-d_{i}(1-\gamma_i)\vv_i$};
\end{tikzpicture}
}
\vspace{-0.7cm}
\caption{Problem geometry for an identifiable system with one \ac{LOS} and one single-bounce \ac{NLOS} path. The dashed gray axes define the global coordinate system $(x,\;y)$, angles of the propagation paths are expressed in the local frame of the \ac{BS} $(x',\;y')$ and \ac{UE} $(x'',\;y'')$ and the notation is introduced in Section \ref{sec:SLAM_position_and_bias}.
}
\label{fig:problem_geometry}
\vspace{-0.3cm}
\end{figure}
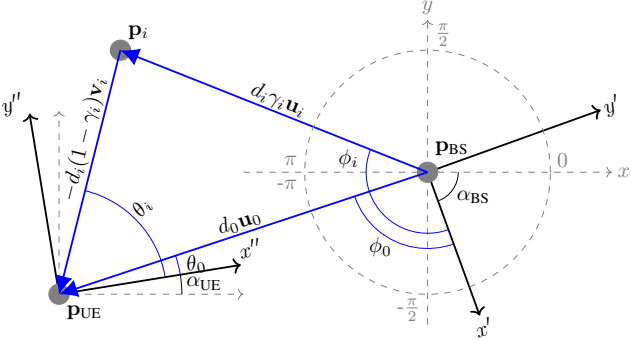

The problem geometry is illustrated in Fig.~\ref{fig:problem_geometry}, considering the \ac{BS} as  transmitter and the \ac{UE} as  receiver. We consider beam-based \ac{OFDM} transmission utilizing beam sweeping at both the \ac{BS} and \ac{UE}. Assuming coarse timing information for obtaining \ac{OFDM} symbols within an applicable time window, under which the channel is assumed constant, the received sample of the $l$th OFDM-symbol and $k$th subcarrier with $m$th \ac{BS} beam and $n$th \ac{UE} beam can be described as~\cite{rastorguevafoi2023}
\begin{align} \label{eq:rx_symbol2}
    y_{k,l}^{m,n}& =\sum_{i=0}^{N}  \xi_i G^m_{\text{BS}}({\phi}_i) G^n_{\text{UE}}({\theta}_i) e^{-\imagunit2\pi k \Delta\!f \tau_i} x_{k,l}^{m,n} + w_{k,l}^{m,n}\!, 
\end{align}
where $\Delta f$ is the subcarrier spacing and $w_{k,l}^{m,n}$ denotes noise after applying the \ac{UE} beam. In addition, $\xi_i$, $\tau_i$, ${\phi}_i$, ${\theta}_i$ are the complex path coefficient, \ac{TOA}, \ac{AOD}, and \ac{AOA} for the $i$th propagation path, respectively. Element $i=0$ is reserved for the \ac{LOS} path (if such is present and detected) and $N$ is the number of \ac{NLOS} propagation paths. To distinguish between these cases, let the hypothesis $\cH=\cH_0$ represent the \ac{LOS} condition, while the alternate hypothesis $\cH=\cH_1$ represents the \ac{NLOS} condition. Furthermore, $G^{m}_{\text{BS}}({\phi}_i) \!\in\! \mathbb{C}$ and $G^{n}_{\text{UE}}({\theta}_i) \!\in\! \mathbb{C}$ are angular responses, containing effects of steering vectors and beamformers,\footnote{More concretely, for example $G^{m}_{\text{BS}}({\phi})=\mathbf{a}_{\text{BS}}^\top(\phi)\mathbf{f}_m$, for BS steering vector $\mathbf{a}_{\text{BS}}^\top(\phi)$ and beamformer $\mathbf{f}_m$.} for the $m$th \ac{BS} beam and  $n$th \ac{UE} beam in respective order. Depending on the knowledge of $G^{m}_{\text{BS}}({\phi}_i)$ and $G^{n}_{\text{UE}}({\theta}_i)$, various methods for estimating the channel parameters exist~\cite{rastorguevafoi2023,shahmansoori2018,cheng2019}. 

\vspace{-2mm}
\subsection{Geometric Measurement Model}

 Let $\vz_i = [{\tau}_i, \, {\phi}_i, \, {\theta}_i]^\top$ denote the channel parameter estimate of the $i$th propagation path, referred to as `measurement' from now on and let $\mathcal{Z}$ and $\mathcal{I}$ denote a set of measurements and associated indices, respectively. Assuming that the measurement noise is zero-mean Gaussian, which is a common assumption in SLAM \cite{kim2020}, the likelihood function is Gaussian 
\begin{equation}\label{eq:slam_likelihood}
    p(\vz_i \mid \vx, \vp_i) = \N(\vz_i \mid \vh(\vx,\vp_i), \vR_i),
\end{equation}
with mean $\vh_i(\vx,\vp_i)$ and covariance $\vR_i$. Building on the geometry in Fig.~\ref{fig:problem_geometry}, the mean of \ac{LOS} and single-bounce \ac{NLOS} paths is given by
\begin{equation}\label{eq:measurement_model}
    \vh(\vx,\vp_i) = \begin{bmatrix}
    d + b_\textrm{UE} \\
    \atan2(-\delta_{1,y},-\delta_{1,x}) - \alpha_\textrm{BS} \\
    \atan2(\delta_{2,y},\delta_{2,x}) - \alpha_\textrm{UE}
    \end{bmatrix}
\end{equation}
and it represents the geometric relationship between the \ac{BS}, \ac{UE} and landmark. For the \ac{LOS} path $(i=0)$, the parameters are defined as: $\vp_i = \emptyset$, $d = \lVert \vp_\textrm{BS} - \vp_\textrm{UE} \rVert/c$ in which $c$ denotes the speed of light and $[\delta_{1,x}, \, \delta_{1,y}]^\top = [\delta_{2,x}, \, \delta_{2,y}]^\top = \vp_\textrm{BS} - \vp_\textrm{UE}$. For the $i$th \ac{NLOS} path, the parameters are defined as: $d = \lVert \vp_\textrm{BS} - \vp_i \rVert/c + \lVert \vp_i - \vp_\textrm{UE} \rVert/c$, $[\delta_{1,x}, \, \delta_{1,y}]^\top = \vp_\textrm{BS} - \vp_i$ and $[\delta_{2,x}, \, \delta_{2,y}]^\top = \vp_i - \vp_\textrm{UE}$. 

\section{Snapshot SLAM}

The procedure to solve the \ac{SLAM} problem depends on $\cH$. We first solve the problem considering $\cH$ to be known and then the method is generalized to detect the \ac{LOS} existence.

\vspace{-3mm}
\subsection{Snapshot SLAM Without Outliers}\label{sec:slam_without_outliers}
 To solve the SLAM problem when there are no outliers, we will first assume that the \ac{UE} orientation is known to provide a conditional estimate, and afterwards estimate the orientation. 
 
\subsubsection{Conditional Position and Clock Bias Estimate}\label{sec:SLAM_position_and_bias}

 Let $\vx_{\alpha_\textrm{UE}} = [\vp_\textrm{UE}^\top, \,  b_\textrm{UE} ]^\top$ denote the partial \ac{UE} state given $\alpha_\textrm{UE}$. As illustrated in Fig.~\ref{fig:problem_geometry}, the \ac{AOD} and \ac{AOA} for a single path can be represented by unit vectors $\vu_i$ and $\vv_i$, given by
\begin{align}
    \vu_i &= \vR(\alpha_\textrm{BS}) \begin{bmatrix}
        \cos(\phi_i) & \sin(\phi_i)
    \end{bmatrix}^\top, \label{eq:bs_unit_vector} \\
        \vv_i &= \vR(\alpha_\textrm{UE}) \begin{bmatrix}
        \cos(\theta_i) & \sin(\theta_i)
    \end{bmatrix}^\top, \label{eq:ue_unit_vector} 
\end{align}
in which $\vR(\alpha)$ is a counterclockwise rotation matrix. Using $\vu_i$ and $\vv_i$, the \ac{UE} position is given by \cite{wen2021}:
\begin{equation}\label{eq:rx_position}
    \vp_\textrm{UE} = \vp_\textrm{BS} + d_i \gamma_i \vu_i - d_i  (1 - \gamma_i) \vv_i
\end{equation}
where $d_i = c (\tau_i - b_\textrm{UE})$ denotes the propagation distance and $\gamma_i \in [0,1]$ is unknown and represents the fraction of the propagation distance along $\vu_i$. 

The position and clock bias can be estimated utilizing the geometric relationship defined in \eqref{eq:rx_position} as follows. First, the expression in \eqref{eq:rx_position} can be rearranged as follows 
\begin{equation}\label{eq:rx_position_rearranged} 
\vH_i \vx_{\alpha_\textrm{UE}} = \vmu_i + \gamma_i d_i \vnu_i
\end{equation}
where $\vH_i = [\vI_{2 \times 2}, \, -c\vv_i]$, $\vmu_i = \vp_\textrm{BS} - c \tau_i \vv_i$ and $\vnu_i = \vu_i  + \vv_i$. Then, we solve for $\gamma_i$ and substitute it back to \eqref{eq:rx_position_rearranged} which yields
\begin{equation}
\vH_i \vx_{\alpha_\textrm{UE}} = \vmu_i + \bar{\vnu}_i^\top (\vH_i \vx_{\alpha_\textrm{UE}} - \vmu_i)\bar{\vnu}_i,
\end{equation}
where $\bar{\vnu}_i = \bar{\vnu}_i/\lVert \bar{\vnu}_i \rVert$. Now, considering all the paths gives the following cost function \cite{ge2023}
\begin{align}
 J(\vx_{\alpha_\textrm{UE}}) &= \sideset{}{_{i \in \mathcal{I}}}\sum  \eta_i J_i(\vx_{\alpha_\textrm{UE}}), \quad \text{where}  \label{eq:cost_function} \\
 J_i(\vx_{\alpha_\textrm{UE}}) &= \lVert \vH_i \vx_{\alpha_\textrm{UE}} - \vmu_i - \bar{\vnu}_i^\top (\vH_i \vx_{\alpha_\textrm{UE}} - \vmu_i)\bar{\vnu}_i \rVert^2 \label{eq:partial_cost}
\end{align}
and $\eta_i = |\hat{\xi}_i|^2$ in which $\hat{\xi}_i$ denotes the complex path coefficient estimate. 
Lastly, a closed-form solution can be obtained by setting the gradient of \eqref{eq:cost_function} to zero and solving for $\vx_{\alpha_\textrm{UE}}$ as
\begin{equation}\label{eq:model_estimate}
    \hat{\vx}_{\alpha_\textrm{UE}} = \left( \sideset{}{_{i \in \mathcal{I}}}\sum \vA_i \right)^{-1} \sideset{}{_{i \in \mathcal{I}}}\sum \vb_i, 
\end{equation}
in which $\vA_i = \eta_i \vH_i^\top \left( \mathbf{I} - \bar{\vnu}_i \bar{\vnu}_i^\top \right) \vH_i$ and $\vb_i = \eta_i \vH_i^\top \left( \mathbf{I} - \bar{\vnu}_i \bar{\vnu}_i^\top \right) \vmu_i$.
The \ac{SLAM} problem is identifiable and all unknowns can be estimated if a sufficient number of \ac{NLOS} paths exist since each \ac{NLOS} path provides three measurements, while being parameterized by two unknowns (i.e., $\vp_i$). With known clock bias either the \ac{LOS} or three \ac{NLOS} are required \cite{mendrzik2019}, whereas one \ac{NLOS} and the \ac{LOS} $(N_\textrm{min} = 2)$ or four \ac{NLOS} $(N_\textrm{min} = 4)$ are needed if the \ac{BS} and \ac{UE} are not synchronized. Fig.~\ref{fig:problem_geometry} illustrates the problem geometry for an identifiable system with a minimum number of paths and it is important to note that for the \ac{LOS} path ($i=0$), we have $\vv_0 = -\vu_0$, $\bar{\vnu}_0 = \mathbf{0}$ and $\gamma_0 = 1$.

\subsubsection{Orientation Estimate}\label{sec:SLAM_orientation}
The procedure to determine the orientation for both $\cH_0$ and $\cH_1$ is described next and the complete state estimate is denoted as $\hat{\vx} = [\hat{\vp}_\textrm{UE}^\top, \, \hat{\alpha}_\textrm{UE}, \,  \hat{b}_\textrm{UE} ]^\top$.
\begin{enumerate}[label=\roman*)]
    \item \textit{LoS condition $\cH_0$} -- From Fig.~\ref{fig:problem_geometry}, we see immediately that the \ac{UE} orientation is given  by 
    \begin{equation}\label{eq:rx_orientation_loS}
        \hat{\alpha}_\textrm{UE} = \atan2({y},{x}),
    \end{equation}
    where $[{x}, \; {y}]^\top = -\vR(-\theta_0) \vu_0$. Hence, the \ac{UE} positioning problem is solved
    by first estimating $\hat{\alpha}_\textrm{UE}$ using \eqref{eq:rx_orientation_loS} and thereafter, $\hat{\vp}_\textrm{UE}^\top$ and $\hat{b}_\textrm{UE}$ are estimated using \eqref{eq:model_estimate}.
    \item \textit{NLoS condition $\cH_1$} -- The problem does not admit a closed-form solution and we must resort to numerical optimization methods as in \cite{wen2021} to obtain the final estimate. Now, the estimate given by \eqref{eq:model_estimate} can be substituted back into \eqref{eq:cost_function} to obtain a cost function over the orientations from which $\hat{\vx}$ can be found by solving
\begin{equation}
    \hat{\alpha}_\textrm{UE} = \underset{{\alpha}_\textrm{UE}\in \mathcal{A}}{\textrm{min} \;} J(\hat{\vx}_{\alpha_\textrm{UE}}),
\end{equation} 
where $\mathcal{A}$ denotes a grid of possible \ac{UE} orientations. 
\end{enumerate}

\subsubsection{Landmark Estimate}

Given $\hat{\vx}$, the landmark locations can be estimated by solving a nonlinear optimization problem for every \ac{NLOS} propagation path $i=1,2, ..., N$, independently. The optimization problem is defined as
\begin{equation}\label{eq:optimization_problem}
    \hat{\vp}_i = \underset{\vp_i}{\textrm{min} \;}  (\vz_i - \vh(\vx,\vp_i))^\top \vR_i^{-1} (\vz_i - \vh(\vx,\vp_i)),
\end{equation}
where $\vh(\vx,\vp_i)$ and $\vR_i$ are given in \eqref{eq:slam_likelihood}. The optimization problem can be solved using the Gauss-Newton algorithm \cite{boyd_vandenberghe_2004} which is initialized using $\vp_i = (\vp_{\textrm{BS},i} + \vp_{\textrm{UE},i})/2$ where $\vp_{\textrm{BS},i}$ and $\vp_{\textrm{UE},i}$ are vectors that span towards $\vp_i$ from the \ac{BS}, $\vp_{\textrm{BS},i} = \vp_\textrm{BS} + \hat{d}_i \hat{\gamma}_i \hat{\vu}_i$ and \ac{UE}, $\vp_{\textrm{UE},i} = \hat{\vp}_\textrm{UE} + \hat{d}_i (1 - \hat{\gamma}_i) \hat{\vv}_i$, respectively, and parameters $\hat{\vu}_i$, $\hat{\vv}_i$, $\hat{d}_i$ and $\hat{\gamma}_i$ are computed by plugging $\hat{\vx}$ into \eqref{eq:bs_unit_vector}, \eqref{eq:ue_unit_vector} and \eqref{eq:rx_position_rearranged}.

\vspace{-6mm}
\subsection{Robust Snapshot SLAM}

\setlength{\textfloatsep}{2mm}

The \ac{SLAM} algorithm described in the previous section is sensitive to measurement outliers. In the worst case, measurements that cannot be expressed using \eqref{eq:rx_position} cause the solution to be very inaccurate. In the following, a robust \ac{SLAM} approach which is inspired by the \ac{RANSAC} algorithm \cite{fischler1981} is described. The aim of the robust \ac{SLAM} algorithm is to find a subset of $\mathcal{Z}$ that does not contain outliers and which is then used to estimate $\hat{\vx}$. Instead of randomly sampling a subset of $\mathcal{Z}$ as in \ac{RANSAC} to determine $\hat{\vx}$, we can perform an exhaustive search over all possible combinations instead, since $\lvert \mathcal{Z} \rvert$ is typically quite low. In the context of \ac{RANSAC}, minimal solvers that compute solutions from a minimal amount of data, $N_\textrm{min}$ in our case, are preferable because it increases the likelihood of using a set that has no outliers. Let, with a slight abuse of notation, 
$\mathcal{J}_\textrm{all}$ denote the set of indices of $\mathcal{Z}$ and 
$\mathcal{J} \in \mathbb{Z}^{L \times N_\textrm{min}}$ the set of all combinations, where $L$ is the total number of combinations 
\begin{equation}\label{eq:number_of_combinations}
L = \frac{(\vert \mathcal{Z} \rvert - N_\textrm{LoS})!}{(\vert \mathcal{Z} \rvert - N_\textrm{LoS} - N_\textrm{NLoS})! \, N_\textrm{NLoS}!},
\end{equation}
in which $N_\textrm{LoS}$ and $N_\textrm{NLoS}$ denote the minimum number of \ac{LOS} and \ac{NLOS} paths required to solve the \ac{SLAM} problem.\footnote{Example for $\lvert \mathcal{Z} \rvert = 4$:   
\begin{enumerate*}[label=(\roman*)]
  \item \emph{\ac{LOS} condition} $\mathcal{H}_0$: $\mathcal{J}_\textrm{all} = \lbrace 0,1,2,3  \rbrace$, $N_\textrm{min}=2$, $N_\textrm{LoS}=1$, $N_\textrm{NLoS}=1$, $L=3$ and $\mathcal{J}  = \lbrace \lbrace0,1\rbrace, \lbrace0,2\rbrace, \lbrace0,3\rbrace \rbrace$;
  \item \emph{\ac{NLOS} condition} $\mathcal{H}_1$: $\mathcal{J}_\textrm{all}  = \lbrace 1,2,3,4  \rbrace$, $N_\textrm{min}=4$, $N_\textrm{LoS}=0$, $N_\textrm{NLoS}=4$, $L=1$ and $\mathcal{J} = \lbrace 1,2,3,4 \rbrace$.
\end{enumerate*}
To note, $\mathcal{I}$ in Section \ref{sec:slam_without_outliers} corresponds to $\mathcal{J}_\textrm{all}$.}

\begin{algorithm}[t]
\caption{Robust snapshot SLAM algorithm}
\label{alg:robust_SLAM_algorithm}
\begin{algorithmic}[1]
\small
\Require Combinations $\mathcal{J} \in \mathbb{Z}^{L \times N_\textrm{min}}$, LoS/NLoS condition $\cH$, $M$,  $\mathcal{Z}$
\Ensure Cost matrix $\mathbf{C} \in \mathbb{R}^{L \times \lvert \mathcal{A} \rvert}$
\State \textbf{if} $\cH=\mathcal{H}_0$ \textbf{then} $\mathcal{A}=\{ 0\}$ 
\textbf{else} $\mathcal{A} = \text{linspace}(-\pi,\pi,M)$
    \For{$m = 1:\lvert \mathcal{A} \rvert$}
    \For{$l = 1:L$}
         \State \textbf{if} $\cH=\mathcal{H}_0$ \textbf{then} $\alpha_\textrm{UE}\leftarrow$  \eqref{eq:rx_orientation_loS} \textbf{else}
          $\alpha_\textrm{UE} \leftarrow \lbrace \mathcal{A} \rbrace_m$ 
          \State Set $C_{l,m} = \infty$
         \State Set $\mathcal{I} \leftarrow \lbrace \mathcal{J} \rbrace_l$ and compute $\hat{\vx}_{\alpha_\textrm{UE}}$ using \eqref{eq:model_estimate} based on $\mathcal{I}$
              \State Set $\mathcal{I}_\textrm{inliers} = \lbrace i \in \mathcal{J}_\textrm{all} \mid J_i(\hat{\vx}_{\alpha_\textrm{UE}}) \leq T_\varepsilon \rbrace$ 
              \State Set $\mathcal{I}_\textrm{outliers} = \mathcal{J}_\textrm{all} \setminus \mathcal{I}_\textrm{inliers}$

         \State 
         Recompute $\hat{\vx}_{\alpha_\textrm{UE}}$ using \eqref{eq:model_estimate}  based on $\mathcal{I}_\textrm{inliers}$
         \If{\texttt{is\_feasible$(\hat{\vx}_{\alpha_\textrm{UE}},\mathcal{I}_\textrm{inliers},\mathcal{Z})$}}
             \vspace{-1mm}
             \begin{equation}
                C_{l,m}= \underbrace{\sum_{i \in \mathcal{I}_\textrm{inliers}} \eta_i J_i(\hat{\vx}_{\alpha_\textrm{UE}})}_{C_{\text{inlier}}} +  \underbrace{\sum_{i \in \mathcal{I}_\textrm{outliers}} \eta_i T_\varepsilon}_{C_{\text{outlier}}}
             \end{equation}
             \vspace{-2mm}
         \EndIf
\EndFor 
\EndFor
\State \textbf{function} \texttt{is\_feasible$({\vx}_{\alpha_\textrm{UE}},\mathcal{I},\mathcal{Z})$}
\vspace{-1mm}
\begin{subequations}\label{eq:feasibility_constraints}
\begin{flalign}
    \quad & j = \argmin_{i \in \mathcal{I}} \, \lbrace \tau_i \rbrace & \nonumber \\ 
    \quad &\textbf{if } \lvert \mathcal{I} \rvert < N_\textrm{min}, \textbf{ return false} & \label{eq:constraint0} \\
    \quad &\textbf{if } \tau_j - {b}_\textrm{UE} < 0, \textbf{ return false} & \label{eq:constraint1} \\
    \quad &\textbf{if } \neg (0 \leq \gamma_i \leq 1 \lor \lVert \vnu_j \rVert^2 \leq T_{\vnu}), \textbf{ return false} & \label{eq:constraint2} \\
    \quad &\textbf{if } \neg (0 \leq \gamma_i \leq 1), \quad i \in \mathcal{I} \setminus \lbrace j \rbrace, \textbf{ return false} & \label{eq:constraint3} \\
    \quad &\textbf{return true} &  \nonumber
\end{flalign}
\end{subequations}
\vspace{-5mm}
\end{algorithmic}
\end{algorithm}

The robust snapshot SLAM problem is defined as follows. For each set $\mathcal{I}\in \mathcal{J}$ and  each value of $\alpha_{\text{UE}}\in \mathcal{A}$ (under $\cH_1$), the corresponding  cost is $\sum_{i \in \mathcal{I}}  \eta_i J_i(\hat{\vx}_{\alpha_\textrm{UE}})$. According to the principles of RANSAC, from the solution $\hat{\vx}_{\alpha_\textrm{UE}}$, the measurements are partitioned into a set of inliers and a set of outliers. A final cost is then computed, denoted by $C(\alpha_{\text{UE}},\mathcal{I})=C_{\text{inlier}}(\alpha_{\text{UE}},\mathcal{I})+C_{\text{outlier}}(\alpha_{\text{UE}},\mathcal{I})$. The penalty $C_{\text{outlier}}(\alpha_{\text{UE}},\mathcal{I})$ is added to ensure that the solution favors estimates obtained with as many inliers as possible.   
Finally, the optimal solution is found as $\min_{\alpha_{\text{UE}},\mathcal{I}}C(\alpha_{\text{UE}},\mathcal{I})$. The complete method is detailed in Algorithm~\ref{alg:robust_SLAM_algorithm}, including the 
computation of the inlier and outlier cost, as well as the search over minimal index sets $\mathcal{I}$ (indexed by $l$) and UE rotations $\alpha_{\text{UE}}$ (indexed by $m$).

To ensure that the solution is physically meaningful, fundamental feasibility constraints are verified for each solution, given in \eqref{eq:feasibility_constraints}. First, the number of inliers should be sufficient to compute $\hat{\vx}_{\alpha_\textrm{UE}}$ as stated in \eqref{eq:constraint0}. Second, the unbiased delay of the shortest path has to be positive as given in \eqref{eq:constraint1}. Third, either the shortest path has to follow \eqref{eq:rx_position} or unit vectors of the shortest path must be opposite and nearly parallel as given in \eqref{eq:constraint2} in which $T_{\vnu}$ is a threshold. Fourth, the constraint in \eqref{eq:constraint3} comes from the definition of $\vp_\textrm{UE}$ in \eqref{eq:rx_position}.

\begin{figure*}
\centerline{
\subfigure[Example solution]{\includegraphics[trim={0.2cm 0.0cm 1.2cm 0.8cm},clip=true,width=6cm]{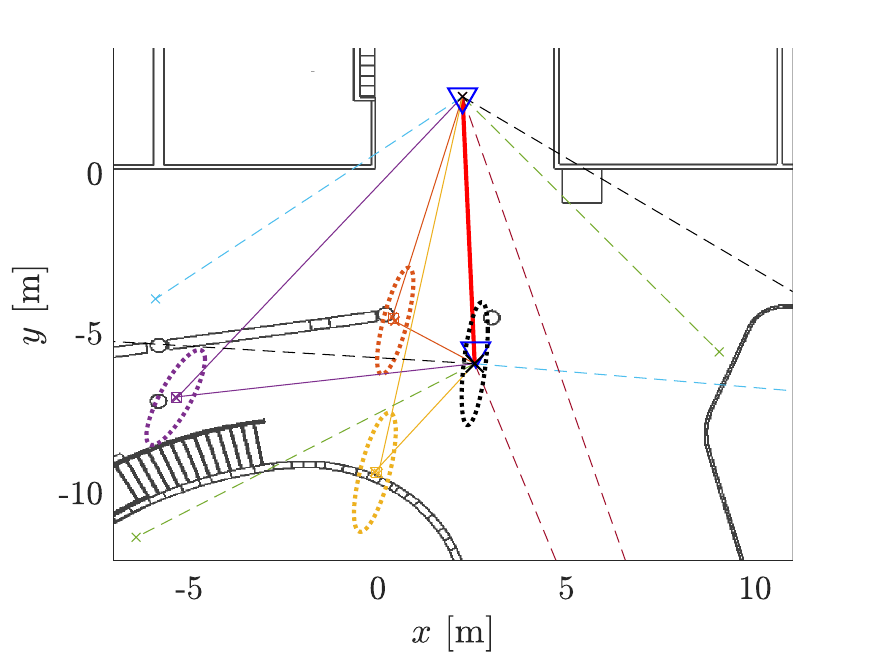}
\label{fig:SLAM_example}}
\hfil
\subfigure[Proposed method]{\includegraphics[trim={0.2cm 0.0cm 1.2cm 0.8cm},clip=true,width=6cm]{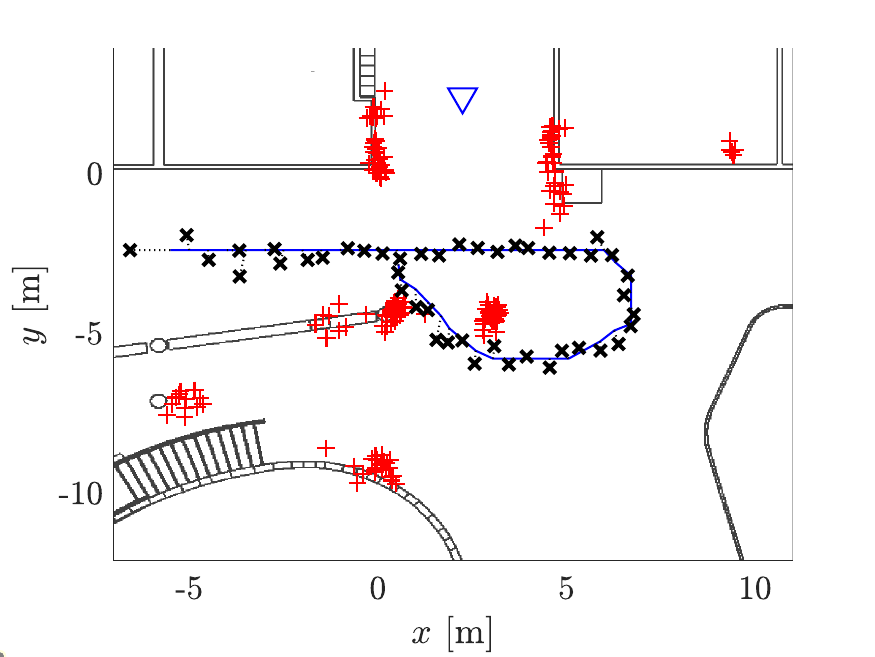}
\label{fig:SLAM_performance}}
\hfil
\subfigure[Benchmark method w/o outliers \cite{wen2021}]{\includegraphics[trim={0.2cm 0.0cm 1.2cm 0.8cm},clip=true,width=6cm]{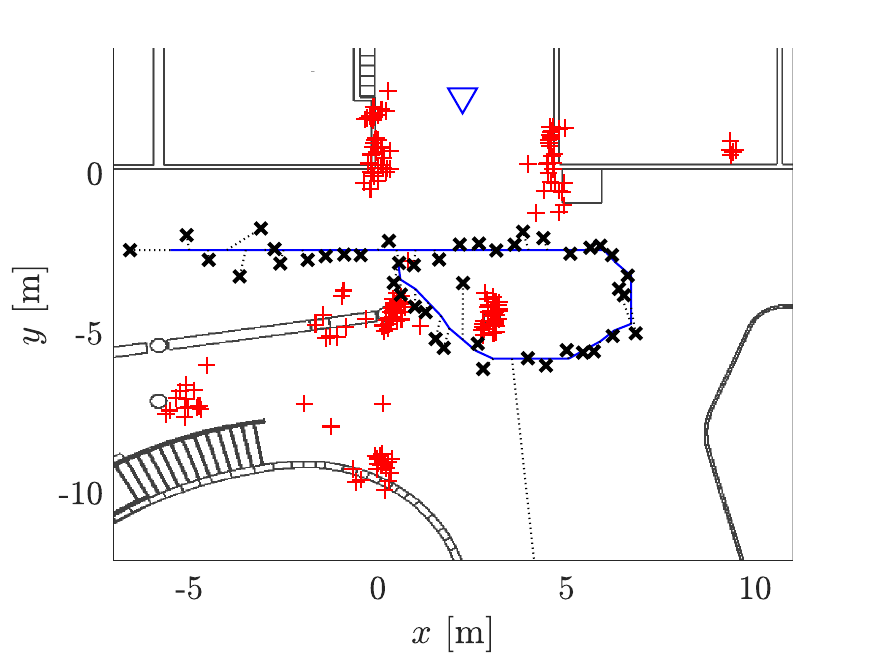}
\label{fig:SLAM_performance_benchmark}}
}
\vspace{-.2cm}
\caption{Example SLAM performance using the proposed and benchmark methods. In (a), the obtained snapshot SLAM solution using the proposed method in one measurement position. In the figure, $\vp_\textrm{BS}$ and $\vp_\textrm{UE}$ illustrated with (\protect\bluetriangle), $\hat{\vp}_\textrm{UE}$ with (\protect\blackcross), LoS path using (\protect\redline), single-bounce NLoS paths with solid lines, propagation paths that are identified as outliers shown using dashed lines (plotted by plugging $\gamma_i = 0.5$ in to the single-bounce model in \eqref{eq:rx_position}) and the dotted ellipses illustrate the one standard deviation uncertainty ellipse of the UE and landmark position estimates. In (b) and (c), $\hat{\vp}_\textrm{UE}$ and $\hat{\vp}_i$'s for all $45$ measurement positions are shown. The legend of the figures is: $\vp_\textrm{BS}$ (\protect\bluetriangle), $\vp_\textrm{UE}$ (\protect\blueline), $\hat{\vp}_\textrm{UE}$ (\protect\blackcross) and $\hat{\vp}_i$'s (\protect\redplus).}
\label{fig:experimental_results}
\vspace{-.4cm}
\end{figure*}

\vspace{-3mm}
\subsection{Extension to LoS Detection and Mixed LoS/NLoS}\label{sec:unknown_los_scenario}

Thus far, $\cH$ has been assumed known and in the following, the proposed robust snapshot \ac{SLAM} algorithm is generalized to detect $\cH_0$ and solve the problem in mixed \ac{LOS}/\ac{NLOS} conditions. In this paper, the \ac{SLAM} problem is solved in mixed \ac{LOS}/\ac{NLOS} conditions by first assuming that $\cH_0$ is true and using a candidate \ac{LOS} path. Thereafter, a statistical test is performed to validate the obtained solution and if the validation test fails, the \ac{SLAM} problem is solved again assuming $\cH_1$. It is to be noted that assuming $\cH_1$ is always possible, but may lead to performance degradation under $\cH_0$ as presented in Section \ref{sec:los_detection_accuracy}. First of all, the \ac{LOS} candidate is the path with smallest delay, $j = \argmin_{i \in \mathcal{J}_\textrm{all}} \, \lbrace \tau_i\rbrace$, after which the estimate $\hat{\vx}$ is obtained using Algorithm~\ref{alg:robust_SLAM_algorithm}. Second, we model the \ac{LOS} signal strength in decibels, $\eta_{j}^\textrm{dB} = 10\log_{10}(\eta_{j})$, using the log-distance path loss model \cite{smulders2009}:
\begin{equation}\label{eq:path_loss_model}
    \eta_{j}^\textrm{dB} \sim \N(f(\vp_\textrm{UE}),\sigma^2), 
\end{equation}
where $f(\vp_\textrm{UE}) = L_0 + 10 \, \zeta \log_{10}(\lVert \vp_\textrm{BS} - \vp_\textrm{UE} \rVert)$, $L_0$ is the reference path loss at a distance of one meter and $\zeta$ the path loss coefficient. The model parameters ($L_0,\zeta,\sigma^2$) can be learned from training data or derived from existing models (see e.g.  \cite[Table III]{smulders2009}). Third, the utilized statistical test is based on the negative log-likelihood of $\eta_{j}^\textrm{dB}$:
\begin{equation}\label{eq:los_detector}
\frac{1}{2}\left( \log(2\pi\sigma^2) + \frac{1}{\sigma^2}\left(\eta_{j}^\textrm{dB} - f(\hat{\vp}_\textrm{UE}))\right)^2\right) \underset{\cH_0}{\overset{\cH_1}{\gtrless}} T_\textrm{LoS}.
\end{equation}
If $\cH_0$ is false, the \ac{SLAM} problem is re-solved assuming $\cH_1$. 

\setlength{\textfloatsep}{1.7\baselineskip plus 0.2\baselineskip minus 0.5\baselineskip}

\section{Experimental Results}\label{sec:results}
\subsection{Experimental Data and Assumptions}

The performance of the proposed and benchmark SLAM algorithms is evaluated using real-world 60\,GHz measurement data, obtained indoors at Tampere University campus, with floorplan as illustrated in Fig.~\ref{fig:experimental_results}. Altogether $45$ UE locations were measured\textemdash $32$ in \ac{LOS} and $13$ in \ac{NLOS} conditions. Beamformed measurements were obtained using $400$\,MHz transmission bandwidth utilizing 5G NR-specified downlink positioning reference signals. The details of the experimental setup,  hardware and channel estimator are detailed in \cite{rastorguevafoi2023}. 

The proposed \ac{SLAM} algorithm utilizes the following parameters in the evaluations. In \ac{NLOS} conditions,  $1^\circ$ angle resolution ($\lvert \mathcal{A} \rvert = 361$) is used. We utilize the path loss coefficient, $\zeta = 1.7$, and standard deviation, $\sigma = 1.8 \text{ dB}$, of the generic $60$ GHz path loss model  \cite[Table III]{smulders2009} and $L_0 = 13 \text{ dB}$ is calibrated from experimental data. The \ac{LOS} detection threshold in \eqref{eq:los_detector} is $T_\textrm{LoS} = 10.8$ which is computed using the inverse  cumulative distribution function of the $\chi^2$ distribution with tail probability $0.001$. The inlier detection threshold used in Algorithm \ref{alg:robust_SLAM_algorithm} is $T_\varepsilon = 0.1$ and the constraint threshold in \eqref{eq:constraint2} is $T_{\vnu} = 0.1$. The proposed method is benchmarked with respect to \cite{wen2021} that exploits the closed form solution to estimate the \ac{UE} position and clock bias introduced in \cite{ge2023}. The benchmark method always assumes $\mathcal{H}_1$.

\subsection{Results}

\subsubsection{Qualitative Comparison}

 The solution in one example measurement position is illustrated in Fig.~\ref{fig:SLAM_example} using the proposed robust snapshot SLAM algorithm. In the example, the proposed method can correctly resolve the \ac{LOS} and single-bounce \ac{NLOS} paths leading to an accurate \ac{SLAM} solution in which the \ac{UE} position estimate is close to the ground truth and the landmark position estimates closely align with the floor plan of the experimental environment. As illustrated in the figure, the propagation paths that are identified as outliers correspond to three double-bounce propagation paths and one triple-bounce propagation path (dashed green line). In all $45$ measurements positions, the proposed method can accurately identify the propagation paths that are inline with model in \eqref{eq:rx_position} leading to good \ac{SLAM} performance as illustrated in Fig.~\ref{fig:SLAM_performance}. 
 Since the benchmark method cannot inherently deal with outliers, they are removed from the data.
 \footnote{Using the ground truth \ac{UE} state $\vx$ and outlier threshold $T_\textrm{outlier}=3$, $\vz_i$ is labeled an outlier if, $\left(\vz_i - \vh(\vx,\hat{\vp}_i) \right)^\top \vR_i^{-1} \left(\vz_i - \vh(\vx,\hat{\vp}_i) \right) > T_\textrm{outlier}$, in which $\hat{\vp}_i$ is computed using \eqref{eq:optimization_problem}.}
 The \ac{SLAM} performance for the benchmark method is illustrated in Fig.~\ref{fig:SLAM_performance_benchmark} and as shown, the method is clearly not as accurate even without outliers. 
 
 \subsubsection{Quantitative Comparison}

Since the ground truth landmark locations are unknown, which is a common problem in \ac{SLAM} when using real-world experimental data, the quantitative evaluation focuses on the \ac{UE} state. The position, heading and clock bias \acp{RMSE} are tabulated in Table~\ref{table:performance_summary}, and the performance is summarized both with and without outliers. The benchmark method yields poor performance with outliers and the accuracy improves notably without outliers. The proposed method yields equivalent \ac{SLAM} accuracy in both cases and the results demonstrate that the presented outlier removal method is effective in practice. In the last two rows of Table~\ref{table:performance_summary}, the \acp{RMSE} are reported separately in the \ac{LOS} and \ac{NLOS} conditions, and as expected, higher accuracy is achieved when the \ac{LOS} is available. 

\begin{table}[!t]
\renewcommand{\arraystretch}{1.1}
\caption{RMSEs and CPU time of the proposed and benchmark methods.}
\vspace{-.2cm}
\centering
\begin{tabular}{|l|c|c|c|c|}
\hline
Method & Pos. [m] & Head. [deg] & Clk. [ns] & Time [ms]  \\  \hline
Proposed  & $0.36$ & $2.05$ & $1.45$ & $3.02$ \\ 
Benchmark  & $11.12$ & $83.79$ & $46.52$ & $0.96$ \\ \hline
Proposed$^\star$ & $0.36$ & $2.05$ & $1.45$ & $0.51$ \\ 
Benchmark$^\star$ & $1.42$ & $4.19$ & $4.80$ & $0.74$ \\\hline
Proposed$^\dagger$ & $0.29$ & $1.95$ & $1.05$ & $0.02$ \\ 
Proposed$^\ddagger$ & $0.49$ & $2.27$ & $2.13$ & $10.40$ \\ \hline
\multicolumn{5}{l}{$^\star$data w/o outliers; $^\dagger$LoS condition  ($32/45$); $^\ddagger$NLoS condition ($13/45$).} \\
\end{tabular}
\label{table:performance_summary}
\vspace{-.5cm}
\end{table}

\subsubsection{LoS Detection Accuracy}\label{sec:los_detection_accuracy}

\begin{figure}[t!]
\centering
\includegraphics[width=0.9\columnwidth,trim={0.6cm 6.0cm 1.3cm 0.0cm},clip]{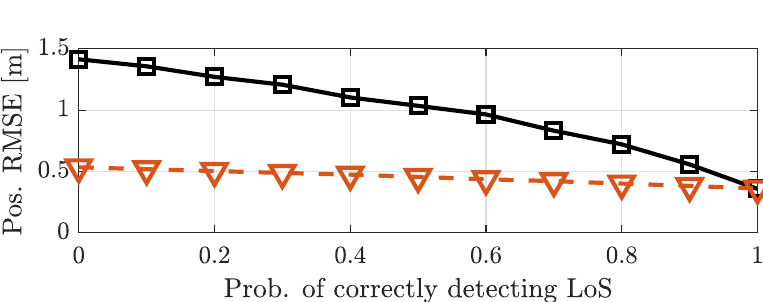}%
\vspace{-.2cm}
\caption{Position RMSE as a function of $P(\mathcal{H}_0 | \text{LoS})$ and the legend is defined as: RMSE for all $45$ measurement positions (\protect\blacksquareline) and RMSE computed without $\vp_\textrm{UE} = [3.56, \, -5.70]^\top$ (\protect\redtriangleline).}
\label{fig:sensitivity_analysis}
\vspace{-.6cm}
\end{figure}

 In the considered experimental setting, the \ac{LOS} detection routine presented in Section \ref{sec:unknown_los_scenario} yields perfect detection accuracy, $P(\mathcal{H}_0 | \text{LoS}) = 1$ and $P(\mathcal{H}_1 | \text{NLoS}) = 1$, but it is to be noted that perfect accuracy is not mandatory. To further investigate \ac{SLAM} performance with an imperfect \ac{LOS} detector, we synthetically vary the probability of correctly detecting the \ac{LOS}, $P(\mathcal{H}_0 | \text{LoS}) \in [0, \, 1]$, and perform $1000$ Monte Carlo simulations with each value of $P(\mathcal{H}_0 | \text{LoS})$. The result is illustrated in Fig.~\ref{fig:sensitivity_analysis}. Since the estimate at $\vp_\textrm{UE} = [3.56, \, -5.70]^\top$ happens to be very inaccurate if the \ac{LOS} is misdetected, the \ac{RMSE} is computed with and without this unfavorable position. Interestingly when $P(\mathcal{H}_0 | \text{LoS}) = 0$, the problem is solved assuming \ac{NLOS} only and the performance is equivalent to the benchmark method without outliers, indicating the proposed method is robust to outliers even without the \ac{LOS} detector. Furthermore, as $P(\mathcal{H}_0 | \text{LoS})$ increases, the \ac{RMSE} decreases which suggests that it is beneficial to solve the problem using the \ac{LOS} if it exists. The main benefit of using the \ac{LOS} is that the \acp{AOD} and \acp{AOA} define the entire geometry up to a scaling since the \ac{UE} orientation is solved in closed-form and the scaling is determined by the clock bias. In addition, the number of degrees of freedom is lower. Thus, the problem is easier to solve and identifiability of the model is improved leading to increased robustness and enhanced \ac{SLAM} accuracy when constraints imposed by the \ac{LOS} are exploited.

\subsubsection{Computational Complexity}
  
The computational complexity of the algorithms are given in the last column of Table~\ref{table:performance_summary}, obtained using a Lenovo ThinkPad P1 Gen 2 with a 2.6 GHz Intel i7-9850H CPU and 64 GB of memory. A naive implementation of the proposed and benchmark algorithms scale according to $\mathcal{O}(\lvert \mathcal{A} \rvert \times L)$ and $\mathcal{O}(\lvert \mathcal{A} \rvert)$, respectively. Solving the \ac{SLAM} problem in \ac{LOS} conditions using the proposed method is very efficient since $\lvert \mathcal{A} \rvert = 1$ and $L = \lvert \mathcal{Z} \rvert - 1$. On the other hand, solving the problem is up to three orders of magnitude higher in \ac{NLOS} conditions since $\lvert \mathcal{A} \rvert = 361$ and typically $L$ is much larger. With respect to the benchmark method, the computational complexity of the proposed method is lower/higher in \ac{LOS}/\ac{NLOS} condition and real-time operation with both methods can be easily guaranteed using for example a Matlab MEX-implementations as tabulated in Table~\ref{table:performance_summary}.

\vspace{-1.5mm}
\section{Conclusions}
The paper presented a robust radio \ac{SLAM} algorithm for mixed \ac{LOS}/\ac{NLOS} environments that can estimate the unknown \ac{UE} state, incidence point of single-bounce propagation paths, as well as existence of the \ac{LOS} path, using measurements that are corrupted by outliers. Analysis was carried out using experimental \ac{mmWave} measurements and the results demonstrated the importance of detecting the outliers and existence of the \ac{LOS}. The newly proposed radio \ac{SLAM} algorithm admits numerous possibilities of future research into other \ac{mmWave} localization and sensing approaches or enhancements via filtering and smoothing.



\end{document}